\documentclass[11pt,twoside]{article}
 
\usepackage{asp2006}
\usepackage{epsf}
\usepackage{psfig}
\usepackage{lscape}
 
\begin{document}

\title{Tremaine-Weinberg integrals for gas flows in double bars}

\author{Witold Maciejewski}
\affil{Astrophysics Research Institute, Liverpool John Moores University, 
Twelve Quays House, Egerton Wharf, Birkenhead CH42 1LD}

\author{Hannah Singh}
\affil{Astrophysics, University of Oxford, Keble Road, Oxford OX1 3RH}

\begin{abstract}
We report on our attempts to achieve a nearly steady-state gas flow in
hydrodynamical simulations of doubly barred galaxies. After exploring the
parameter space, we construct two models, for which we evaluate the 
photometric and the kinematic integrals, present in the Tremaine-Weinberg 
method, in search of observational signatures of two rotating patterns.
We show that such signatures are often present, but a direct fit to data 
points is likely to return incorrect pattern speeds. However, for a particular
distribution of the tracer, presented here, the values of the pattern speeds 
can be retrieved reliably even with the direct fit.
\end{abstract}

\section{Introduction}
Bars within bars are common in disc galaxies -- up to 30\% of early-type
barred galaxies contain them (Erwin \& Sparke 2002). It is generally expected 
that they play a role in the feeding of active galactic nuclei, though
the direct theoretical confirmation is difficult, as the interaction between 
the bars considerably narrows the range of possible systems (Maciejewski \& 
Sparke 2000, hereafter MS00; Maciejewski \& Athanassoula 2008). Corsini, 
Debattista, \& Aguerri (2003) demonstrated that in NGC 2903 the two bars rotate
with two different pattern speeds. Maciejewski (2006) and Merrifield, Rand,
\& Meidt (2006) proposed extensions to 
the Tremaine-Weinberg method (Tremaine \& Weinberg 1984) in order
to derive multiple pattern speeds. There were also attempts to derive them
from direct fitting of straight lines to the Tremaine-Weinberg (TW) 
integrals. Here we analyse in detail the behaviour of the TW integrals in 
hydrodynamical models of gas flow in double bars. Since only one dynamically 
plausible model of double bars has been constructed so far (MS00), we 
approach the problem from another direction here: by changing the parameters 
of the two bars, we search for steady-state flows, which may indicate 
preferred systems.

\begin{figure}[t]
\plotone{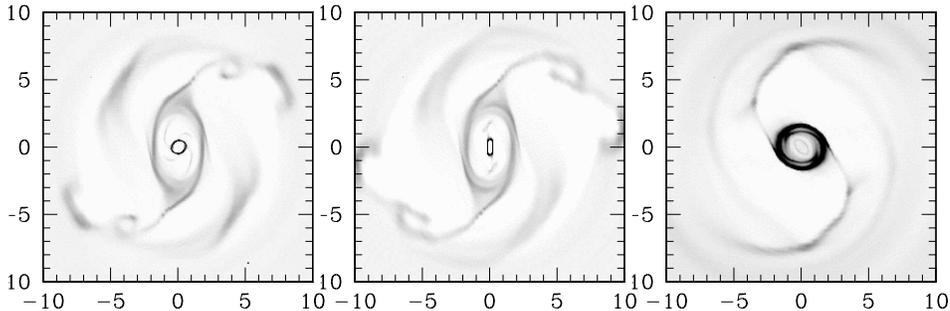}
\vspace{-137mm}
\caption{\small Snapshots of gas density in two models of gas flow in double 
bars: Setup 1 at 955 Myr (left panel) and at 980 Myr (central panel); Setup 2 
at 1445 Myr (right panel). Darker shading indicates higher density. Both bars 
rotate counterclockwise and the outer bar is vertical on the plots. Units 
on the axes are in kpc.}
\end{figure}

\section{Code and models}
We used the grid-based Eulerian hydrodynamical code CMHOG written by James M. 
Stone and adopted to the polar grid in two dimensions by Piner, Stone, \& 
Teuben (1995). The code implements the piecewise parabolic method (PPM), 
it uses the 
isothermal equation of state and it does not account for self-gravity in gas.
From the set of models by Regan \& Teuben (2003, hereafter RT03) of gas flow 
in a single bar, we picked the ones in which the flow is likely to become a
steady state after the secondary bar is introduced. This is likely if in the 
single bar the gas settles on an oval around that bar, with no features in 
the centre, where the secondary bar is going to be placed. The models in RT03,
which consider cold gas (5 km s$^{-1}\,$ speed of sound), are particularly 
suitable, since for warmer gas a nuclear spiral develops in the centre 
(Maciejewski 2004). 

The model from the bottom-right panel in fig.3 of RT03 (axial ratio of the bar 
$a/b=2.0$, bar quadrupole moment $Q_M = 12.5 \times 10^{10}$ M$_{\odot}$ 
kpc$^{-2}$, central mass concentration $\rho_c=1.0 \times 10^{10}$ M$_{\odot}$ 
kpc$^{-3}$) is the best case
of the oval flow around the bar within which the secondary bar can be placed
(Setup 1). We also searched for models with the nuclear ring sufficiently large
that the flow patterns induced by the secondary bar could fit within it (Setup 
2). As the starting point for that setup, we used the models from the top row 
of fig.3 in RT03 ($a/b=2.0$, $\rho_c=3.5$), but in order to increase the size 
of the nuclear ring, we slowed down the rotation of the bar by increasing its 
Lagrangian radius $r_L$ (measured at the L1 point) from 6 kpc in RT03 to 11 
kpc. In total we constructed 21 models by varying $a/b$ between 1.8 and 2.5, 
$Q_M$ between 4.5 and 12.5, and $\rho_c$ between 1.0 and 4.8, in the units 
above. Five out of these 21 models reproduce well the two desired setups 
described above: gas flows for Setup 1 appear to reach steady-state within 
the oval flow, but not outside it, while gas flows for Setup 2 are steady-state
throughout, although strong shocks persist along the bar.

For the introduction of the secondary bar we used the parameters of the five 
models above. We constructed 5 models of gas flow in double bars, out of which
two closest to the steady-state have the following parameters for Setup 1 (2):
$a/b=1.8 (2.0)$, $Q_M = 12.5 (5.0)$, $\rho_c=1.0 (3.5)$, $r_L=6 (11)$, in the
units above, with the parameters of the inner bar close to those from Model 2 
in MS00: the ratio of the semi-major axes 0.2, the mass ratio 0.10 (0.15), 
the small bar axial ratio 2, and its pattern speed 110 km s$^{-1}$ kpc$^{-1}$.
Snapshots of gas density in these two
models are showed in Fig.1. The flow in the region of the inner bar in Setup 1
is far from a steady state. A gaseous oval following that bar is flattened
when the bars are aligned, and nearly circular at bars perpendicular. This is
inconsistent with the orbital structure found in Model 2 of MS00. However,
the low central mass concentration in Setup 1 may imply that the $x_2$ orbital 
family that gives rise to the orbits supporting the inner bar in Model 2
from MS00 is absent there. Throughout both bars, the gas flow in Setup 1 is 
likely to follow orbits that originate from the $x_1$ orbits in a single bar.
In Setup 2, most of the gas accumulates in the region dominated by the inner
bar and the flow is much closer to the steady-state than in Setup 1 or in
Model 2 of MS00 (see Maciejewski et al.~2002).

\begin{figure}[t]
\plotone{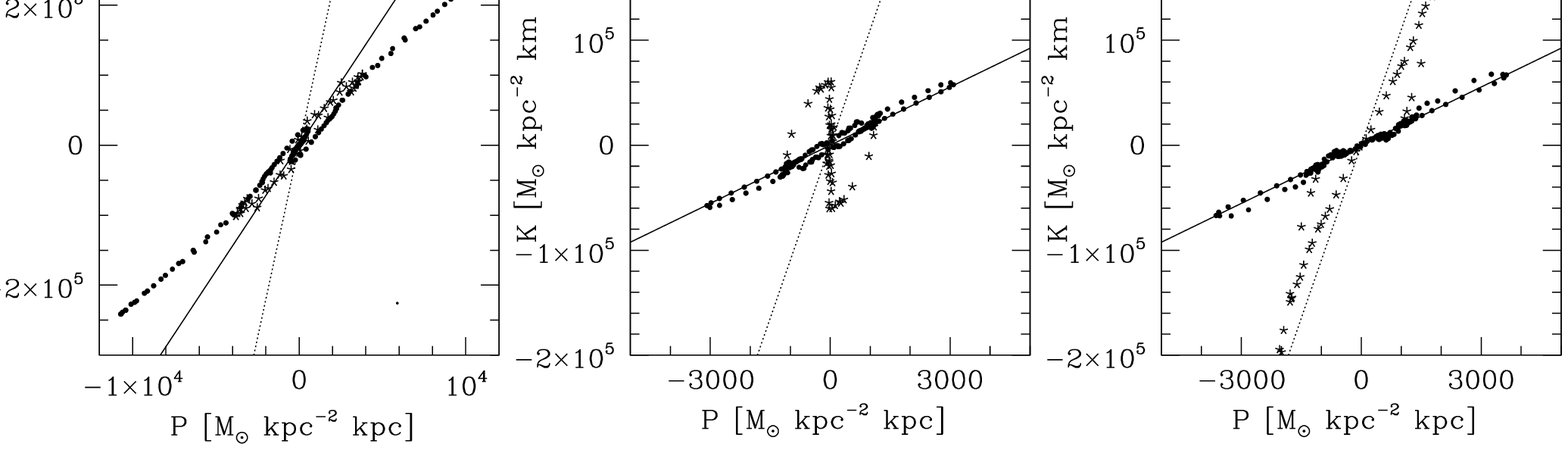}
\vspace{-134mm}
\caption{\small The photometric TW integral $P$, on the horizontal axis, 
plotted against the kinematic integral $K$, on the vertical axis, both 
measured along the slits parallel to the line of nodes. Stars mark the 
integrals for the slits offset by less than 2 kpc from the line of nodes 
(horizontal in Fig.1), while filled circles correspond to the slits further 
out. The slope
of the solid (dotted) line is equal to the imposed pattern speed of the outer
(inner) bar. The diagrams are constructed for Setup 1 at 835 Myr (left panel),
and Setup 2 at 755 Myr (central panel) and at 1005 Myr (right panel).}
\end{figure}

\section{Tremaine-Weinberg integrals for the models with two bars}
In the two models described above, we followed the gas flows for 3 Gyr, 
recording the hydrodynamical variables (density, velocity) on the grid every 
5 Myr. Two pattern speeds are imposed in our models, and the flow is often 
far from a steady state, hence the assumptions of the Tremaine-Weinberg 
method are violated here. However, even with the assumptions of the
method violated, the TW integrals remain well defined, and the hydrodynamical 
variables that we recorded are sufficient to determine them.

For each setup, we calculated the TW integrals at 600 snapshots in time, and 
searched for a signal of rotating patterns. As our models evolve, we first 
introduce the outer bar, and then the inner bar. In the early evolutionary
stages, with only the outer 
bar present, the relation between the TW integrals is linear, having the slope
consistent with the imposed pattern speed, despite the flow not being
completely settled. Once the secondary bar is introduced in Setup 1, the
flow becomes very unsettled, and for the great majority of snapshots no
relation between the TW integrals is observed. On rare occasions when there
is a linear relation, its slope is inconsistent with the imposed pattern 
speeds, and there is no indication of two distinct rotating patterns -- see 
Fig.2, left panel.

In Setup 2, the modifications of the flow in the outer bar caused by the 
introduction of the inner bar are much smaller than in Setup 1, and two 
steady flows, generated by the two bars, are being established. 
In the majority of snapshots, the points in the diagram for
the TW integrals gather on two distinct lines (Fig.2, central and right
panels). The points for the slits far from the line of nodes, where the outer 
bar dominates the flow, gather on a line whose slope is consistent with the 
imposed pattern speed of the outer bar. However, in the early evolution of the 
model, points for the slits close to the line of nodes gather on a line whose
slope can vary with time, and is clearly inconsistent with the imposed constant
pattern speed of the inner bar (Fig.2, central panel). This changes later in
the run, when both sets of points indicate two correct pattern
speeds at most of the snapshots (Fig.2, right panel).

\section{Discussion and conclusions}
Correct pattern speeds in Setup 2 are recovered only at the late stages of 
evolution, because by then the majority of gas is accumulated in the region 
dominated by the inner bar. Earlier in the run, the slits close to the line of 
nodes sample both flows dominated by the inner and the outer bar. At the end of
the run, the contribution from the outer bar is reduced, so that {\it in the 
slits close to the line of nodes most of the tracer follows the inner bar.} 
This is the only gas distribution for which two pattern speeds can be derived 
with good confidence from a direct fit of straight lines to the TW integrals. 
Otherwise, the points indicating the TW integrals may gather on two lines,
but the slope of the second line may show no relation to the pattern speed of
the inner bar. Moreover, for many systems with two pattern speeds, points 
representing the TW integrals may not gather on any line, as in Setup 1. We
conclude that if there is more than one rotating pattern in a galaxy, the 
direct fits to the TW integrals may provide reliable pattern speeds only for
a very specific distribution of the tracer. Fully comprehensive work on this
subject will be possible once we know the range of systems with two pattern 
speeds that are dynamically possible.

\acknowledgements This work was supported by the Polish 
Committee for Scientific Research as a research project 1 P03D 007 26 in 
the years 2004--2007.

\end{document}